\documentclass{PoS}
\usepackage{lineno}
\usepackage{graphicx}
\usepackage{caption}
\usepackage{subcaption}

\newcommand{\vpt}{\vec{p}_\mathrm{T}}
\newcommand{\mtt}{m_{\mathrm{T}2}}
\newcommand{\mt}{m_{\mathrm{T}}}
\newcommand{\invisible}{\mathrm {invis}}
\newcommand{\visible}{\mathrm{vis}}

\title{Measurement of the top quark mass with the ATLAS detector}

\ShortTitle{Measurement of the top quark mass with the ATLAS detector}

\author{\speaker{Giuseppe Salamanna}\\
        {\rm On behalf of the ATLAS Collaboration}\\
        Queen Mary, University of London\\
        E-mail: \email{Giuseppe.Salamanna@cern.ch}} 

\abstract{An overview is presented of the measurements of the top quark mass performed by the ATLAS experiment at the LHC with an integrated luminosity varying between 35 pb$^{-1}$ and 4.7 fb$^{-1}$. Different techniques are used to measure the top quark mass looking at events in all three signatures: fully-hadronic, lepton+jets and di-leptonic ones. The most precise measurement, using a template method on lepton+jets events, yields a top quark mass of 174.5 $\pm$ 0.6 (stat) $\pm$ 2.3 (syst) GeV. The dominant systematic uncertainties are related to the determination of the $b-$jet energy scale and the modelling of additional radiation accompanying the $t\bar{t}$ pair production.}

\FullConference{36th International Conference on High Energy Physics\\
   4-11 July 2012\\
    Melbourne, Australia}

\begin{document}
\linenumbers

\section{Introduction}
The top quark is the sixth and heaviest of all known quarks. Thanks to its large mass ($m_{top}$, world average value is 173.5 $\pm$ 0.6 (stat) $\pm$ 0.8 (syst) GeV \cite{bib:pdg2012}), it plays an important role in the electro-weak (EW) sector of the Standard Model (SM). In particular, through loop corrections, it affects the value of the W and Z boson masses. Being the only known fermion with a Yukawa coupling close to unity, it also proves to be a primary tool in the investigation of the Higgs boson properties. Its substantial radiative corrections modify the value of the Higgs boson mass with respect to tree level expectations, causing the so-called hierarchy problem. Now that a candidate Higgs boson has been identified at the LHC \cite{bib:atlh,bib:cmsh} and we have direct access to its mass, measurements of $m_{top}$, together with the W boson mass, are motivated by the need to test the internal consistency of the SM. The relationship between the EW parameters can be found, e.g., in \cite{bib:heinemeyer} and updated in \cite{bib:heineweb}. 

It is useful to remind that the results presented in the following are not directly the values of $m_{top}$ used in the EW fits. Being the top quark a coloured object, we do not have direct access to $m_{top}$ in the renormalization scheme at the experimental level. Rather, what we measure is the invariant mass of the products of the top quark decay, a b-hadron and a W boson. With this in mind, we in fact measure an artificial top quark mass injected in the simulations used for the analysis, which is closer to the so-called ``pole mass''. This is then translated into the renormalized mass used for theoretical inputs.    

\section{Collision data and simulated samples}
The measurements presented in this paper are based on different samples of LHC proton-proton collision data, all collected by the ATLAS detector \cite{bib:atlas} in 2011 at a centre-of-mass energy $\sqrt{s}=$ 7 TeV. The results from the lepton+jets sample use an integrated luminosity of 1.04 fb$^{-1}$; the one in the di-leptonic sample makes use of 4.70 fb$^{-1}$, whilst the all-hadronic measurement is performed on a sample of 2.04 fb$^{-1}$. 

On the simulation side, $t\bar{t}$ events are generated using the Next-to-Leading Order (NLO) Monte Carlo (MC) MC@NLO \cite{bib:nlo} with the NLO parton density function sets CTEQ6.6 \cite{bib:cteq} or CT10. The parton showering is modelled using the HERWIG \cite{bib:herwig} programme. The events are simulated using the full GEANT4-based \cite{bib:g4} ATLAS detector simulation. To obtain $m_{top}$ samples are generated with different input mass values ranging from 150 GeV to 200 GeV and processed with the fast version of the ATLAS detector simulation \cite{bib:af2}.
The main backgrounds to the $t\bar{t}$ process are QCD multi-jet events and the production of a W boson in association with jets. Both these processes are estimated directly using collision data, both in terms of rates and shapes. More details on how these are estimated are provided, e.g., in \cite{bib:masstemp}.
Additional backgrounds and generation-related systematic uncertainties are all studied using dedicated simulated samples.  
 
\section{Event selection}
Events in the lepton+jets or di-leptonic channel are selected by a single lepton trigger with a transverse energy threshold of 18 GeV (muon) or 20 GeV (electron). A multi-jet trigger requiring at least 5 jets with transverse momentum ($p_{T}>$30 GeV) is used for the all-hadronic measurement. Electron candidates must have a transverse energy $E_{T}>35$ GeV, a pseudorapidity $|\eta|<$2.5 (but excluding the range 1.37-1.52 in $|\eta|$, where the calorimetry is not instrumented) and be isolated in the calorimetry (the energy not associated to the electron cluster and lying within a cone $\Delta R\footnote{$\Delta R=\sqrt{\Delta\eta^2+\Delta\phi^2}$}=0.2$ of the electron momentum direction should be less than 3.5 GeV). Muons are required to have $p_{T}>20$ GeV and $|\eta|<$2.5 and are selected if isolated in both the calorimeters and at the tracking level (the energy in a cone $\Delta R=0.3$ around the muon in both detectors should not exceed 4 GeV). Jets are reconstructed using the anti-$k_t$ algorithm \cite{bib:antikt} with a radius $R=0.4$ from adjacent energy clusters in the ATLAS calorimetry. Muons closer than $\Delta R=0.3$ to a selected jet are rejected as coming from Heavy-Flavour (HF) decays. For the measurements in the lepton+jets (di-leptonic) sample four (two) jets of $p_T>20$ GeV and $|\eta|<$2.5 are required to select the event. For the fully-hadronic case five jets of $p_T>55$ GeV and $|\eta|<$2.5 are requested, plus a softer jet ($p_T>30$ GeV). 

The results presented in this paper are based on jet calibrations developed on the 2010 and 2011 data, with a generic uncertainty for the energy scale of an inclusive sample (JES); and specific $b$-jet energy scale ($b$-JES) uncertainty. The measurement in the di-leptonic sample also profits from a JES calibration which became available late in 2011. Typical JES uncertainties for the $p_T$ range of $t\bar{t}$ jets are 2-3$\%$ or 1$\%$ with the final calibrations \cite{bib:jetconfnote}. The additional $b$-JES uncertainty is 1 to 2.5$\%$. 

Event-level requirements are also applied on different variables depending on the signature. The missing transverse energy $E_T^{miss}$ is reconstructed from the vector sum of the energy deposits in the calorimetry, calibrated at the electromagnetic scale; plus the muon energy in the calorimetry (if the muon is isolated) and in the tracking volume. For the electron+jets case, a requirement of $E_T^{miss} > 35$ GeV and a transverse reconstructed W boson mass $M_T(W) > 25$ GeV are imposed, to suppress the large background from multi-jet events. For the muon+jets case this becomes: $E_T^{miss} > 20$ GeV and $E_T^{miss} + M_T(W)> 60$ GeV. One of the jets should have been identified as being originated in a $b$ quark decay by a Neural Network based secondary vertex tagger \cite{bib:nntagger}. In the di-leptonic case, two $b$-tags are requested on jets of $p_T>45$ GeV. Also, the scalar sum of transverse momenta of the two selected leptons and all selected jets in the event ($H_T$) should be larger than 130 GeV. Similar quantities are also used in the all-hadronic case, together with a lepton veto.

\section{Template method}
\subsection{Lepton+jets channel}
In order to attain a precision compatible with the best measurements available, techniques are put in place in ATLAS to constrain the JES uncertainty {\it in-situ} using the same $t\bar{t}$ decays from which $m_{top}$ is measured. The first measurement uses templates of the reconstructed top quark mass in the simulation to extract $m_{top}$ in data. The analysis is described in detail in \cite{bib:masstemp}. The sensitivity of the measurement to the determination of the JES is reduced by performing a simultaneous fit of $m_{top}$ and of a {\it global} jet scaling factor (JSF). The quantities looked at are the di-jet and tri-jet invariant masses, that are supposed to come from the W boson and full top quark decays respectively. The JSF is extracted from comparing the predicted di-jet distribution with the one observed in data; it entails JES calibration effects as well as event modelling inaccuracies of the MC with respect to data. By identifying the template that best fits the data out of a two-dimensional set of templates from different $m_{top}$ and JSF values, a top quark mass measurement less sensitive to JES calibrations is obtained. Since the JSF is extracted from the light flavour di-jet pair, this method does not constrain the b-JES uncertainty, which remains the dominant systematic uncertainty. The linearity of the unbinned likelihood fit to the signal and backgrounds with $m_{top}$ is checked using pseudo-experiments. The results of the fit are shown in Figure \ref{fig:templates} and yield $m_{top}$ = 174.5 $\pm$ 0.6 (stat) $\pm$ 2.3 (syst) GeV. Together with the b-JES (1.58 GeV), the most important uncertainties are related to the event modelling in MC: initial and final state radiation (IFSR, 1.01 GeV) and colour reconnection (0.55 GeV). An ongoing effort is aimed at constraining the MC modelling systematic variations directly with ATLAS data. It should be noticed that this effort has already produced a 50$\%$ reduction of the IFSR part since this measurement was performed \cite{bib:jetveto}; the impact of this reduction on $m_{top}$ is being studied. 

This result is cross-checked using a complementary one-dimensional template fit which uses the different approach of fitting a kinematic variable constructed to minimise the sensitivity to JES (ratio of the per-event reconstructed invariant masses of the hadronically decaying top quark and W boson). The results in this case is $m_{top}$ = 174.3 $\pm$ 0.9 (stat) $\pm$ 2.5 (syst) GeV, compatible with the former.

\subsection{All-hadronic channel}
\begin{figure}
  \centering
  \begin{subfigure}[b]{0.5\textwidth}
    \centering
    \includegraphics[width=\textwidth]{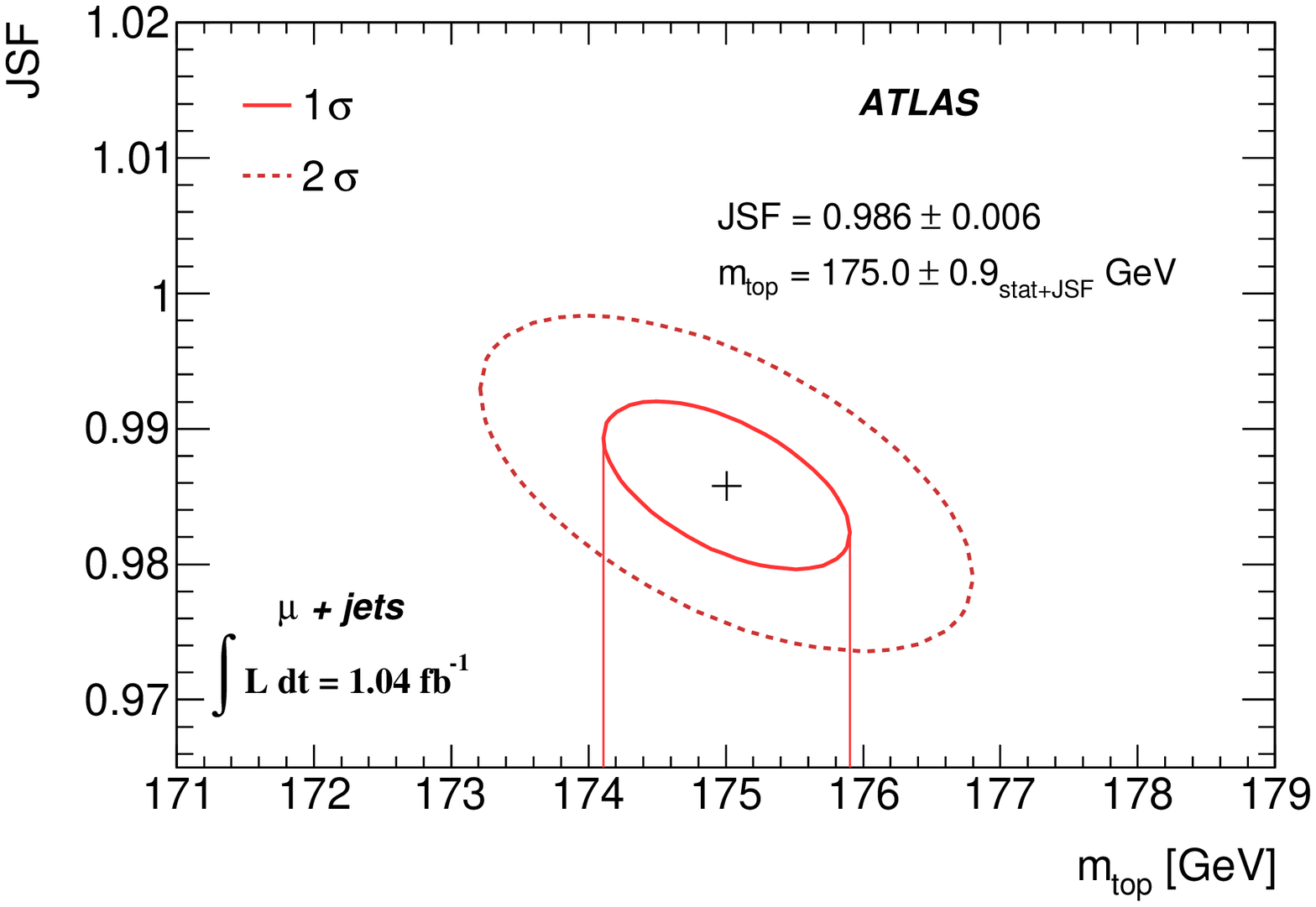}
  \end{subfigure}%
  ~~ 
  \begin{subfigure}[b]{0.5\textwidth}
    \centering
    \includegraphics[width=\textwidth]{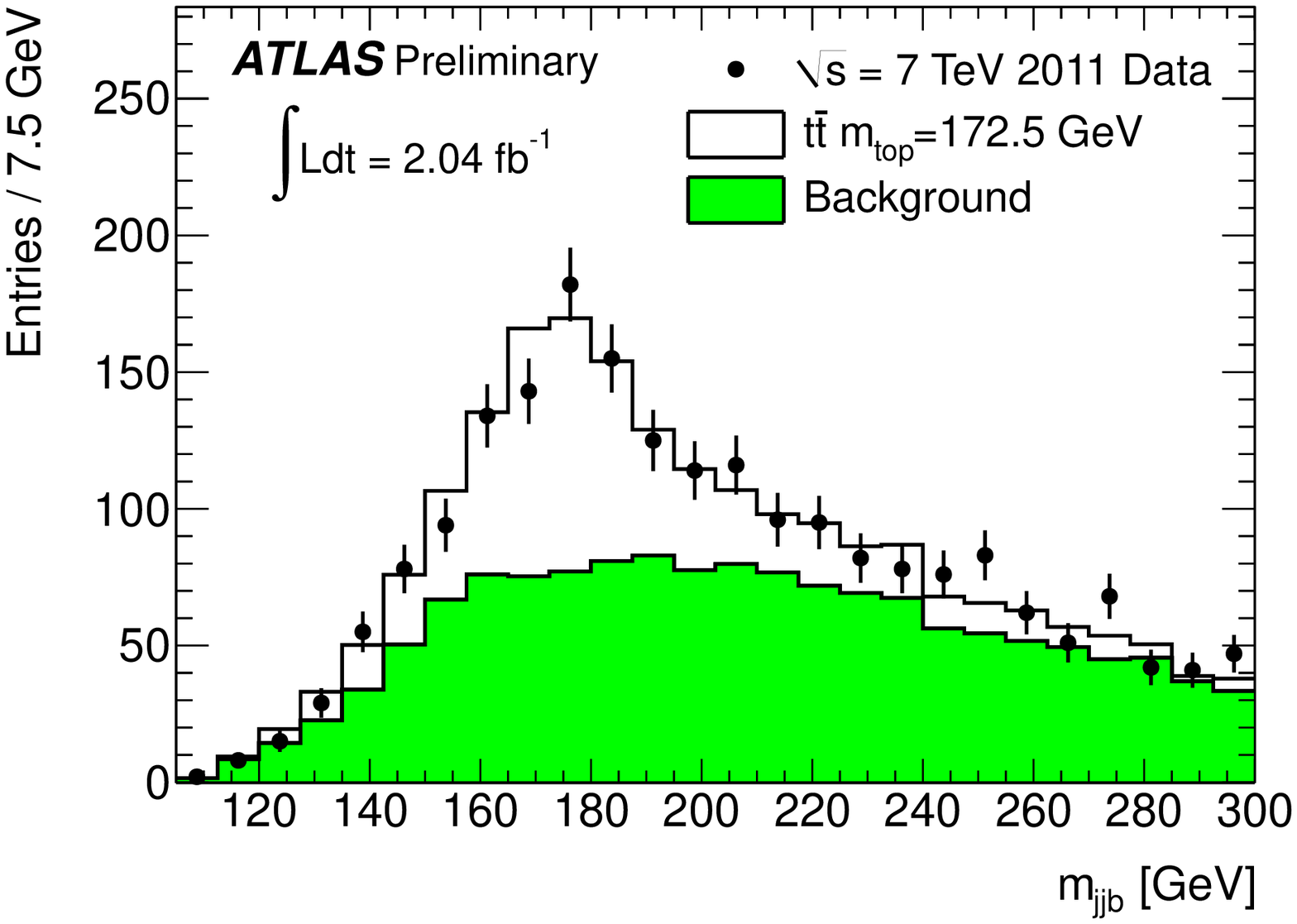}
  \end{subfigure}
  \caption{Left: Correlation of the measured top quark mass and jet energy scale factor with the two-dimensional template method in the $\mu$+jets channel. The ellipses show the 1- and 2-$\sigma$ uncertainties on the two parameters \cite{bib:masstemp}. Right: Distribution of the reconstructed tri-jet invariant mass on the selected events in the all-hadronic channel, with an estimate of the background contribution. \cite{bib:allhad}. \label{fig:templates}} 
\end{figure}

A measurement of $m_{top}$ has also been performed in the all-hadronic $t\bar{t}$ channel, using a similar template technique described in detail in \cite{bib:allhad}. The top quark candidates are identified as those 3-jet combinations which minimize a $\chi^2$ measuring the compatibility of the jets with a particular assignment to the $t\bar{t}$ decay products. Only events whose minimum $\chi^2$ is less than 8 are retained for the mass extraction. The biggest challenge for this measurement lies in a correct estimation of the very large QCD multi-jet background. To this end, a technique called ``event mixing'' is employed. This background is modelled from signal-like events with exactly five jets, to which jets with transverse momentum lower than that of the fifth highest transverse momentum jet from events with six or more jets have been added, so to probe the cross-talk between signal and samples enriched in background. The templates extracted from the samples thus obtained are validated on a sample constructed ad-hoc by looking at events with four hard jets and all of the lowest transverse momentum jets from a signal-like 5-jet event. The resulting reconstructed top quark mass is shown in Figure \ref{fig:templates}. $m_{top}$ has been determined by fitting one-dimensional templates from MC (signal) and data (multijet background) to the reconstructed tri-jet invariant mass distribution using a binned likelihood. The background fraction has been left as a free parameter in the fit. The result is $m_{top}$ = 174.9 $\pm$ 2.1 (stat) $\pm$ 3.8 (syst) GeV.   

\section{Calibration curve method} 
An alternative method is employed to measure $m_{top}$ in the di-leptonic channel, namely in events with one selected electron $e$ and one selected muon $\mu$. The result is documented in \cite{bib:calibc}. In these events the dependence on the knowledge of JES is less pronounced, but it comes at the price of a challenging top candidate reconstruction. The presence of two neutrinos from the leptonic decays of both W bosons doesn't allow to apply kinematic constraints to assign the experimental objects to one of the two top quarks in the event. In order to extract $m_{top}$, the variable $m_{T2}$ is used. This is conceived for final states with more than one undetected particle (like in Supersymmetric searches) \cite{bib:mt2}. It is defined as: 
\begin{equation}
   \label{eqKmttdef}
   \mtt (m_{\invisible})= \min_{\vpt^{\,(1)},\; \vpt^{\,(2)}}
     \biggl\{ \max \Bigl[
       \mt (m_{\invisible}, \vpt^{\,(1)}),
       \mt (m_{\invisible}, \vpt^{\,(2)})
     \Bigr] \biggr\} \, ,
 \end{equation}
where the variables $\vpt^{\,(1)}$ and $\vpt^{\,(2)}$ represent kinematically allowed test values of the invisible particle transverse momenta and
 \begin{equation}
   \label{eqKmtdef}
   \mt (m_{\invisible}, \vpt^{\,(i)}) =
     \sqrt{
       m_{\visible}^{2}+m_{\invisible}^{2}
       + 2(E_{\mathrm{T}}^{\visible}E_{\mathrm{T}}^{\invisible}-\vpt^{\,\visible}
\cdot \vpt^{\,(i)} )
     }\, .
 \end{equation}
Here ``vis'' indicates the visible part of the decay of each of the two top quarks; and ``invis'' stands for the neutrinos. 

\begin{figure}
  \centering
  \begin{subfigure}[b]{0.4\textwidth}
    \centering
    \includegraphics[width=\textwidth]{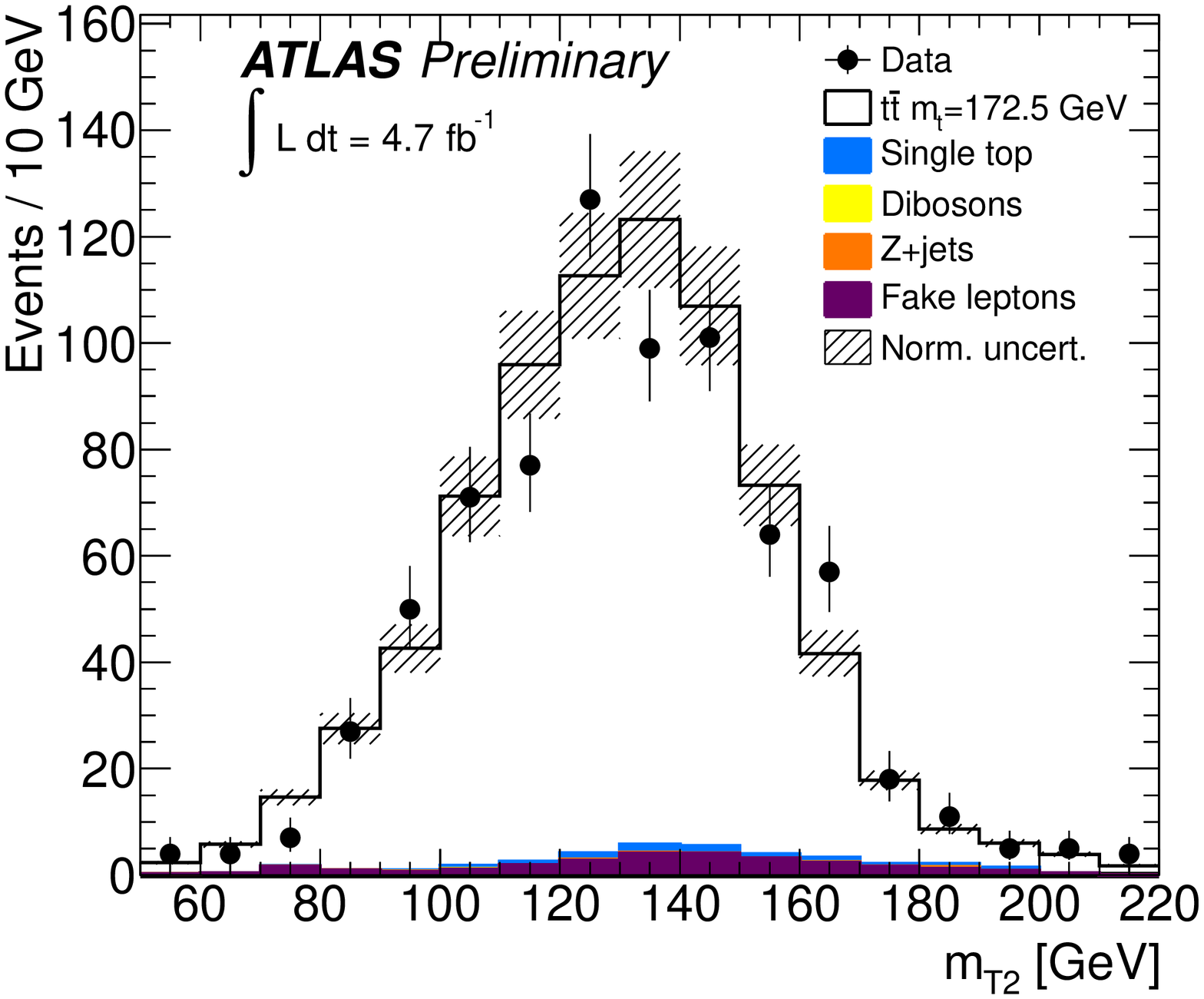}
  \end{subfigure}%
  ~~ 
  \begin{subfigure}[b]{0.4\textwidth}
    \centering
    \includegraphics[width=\textwidth]{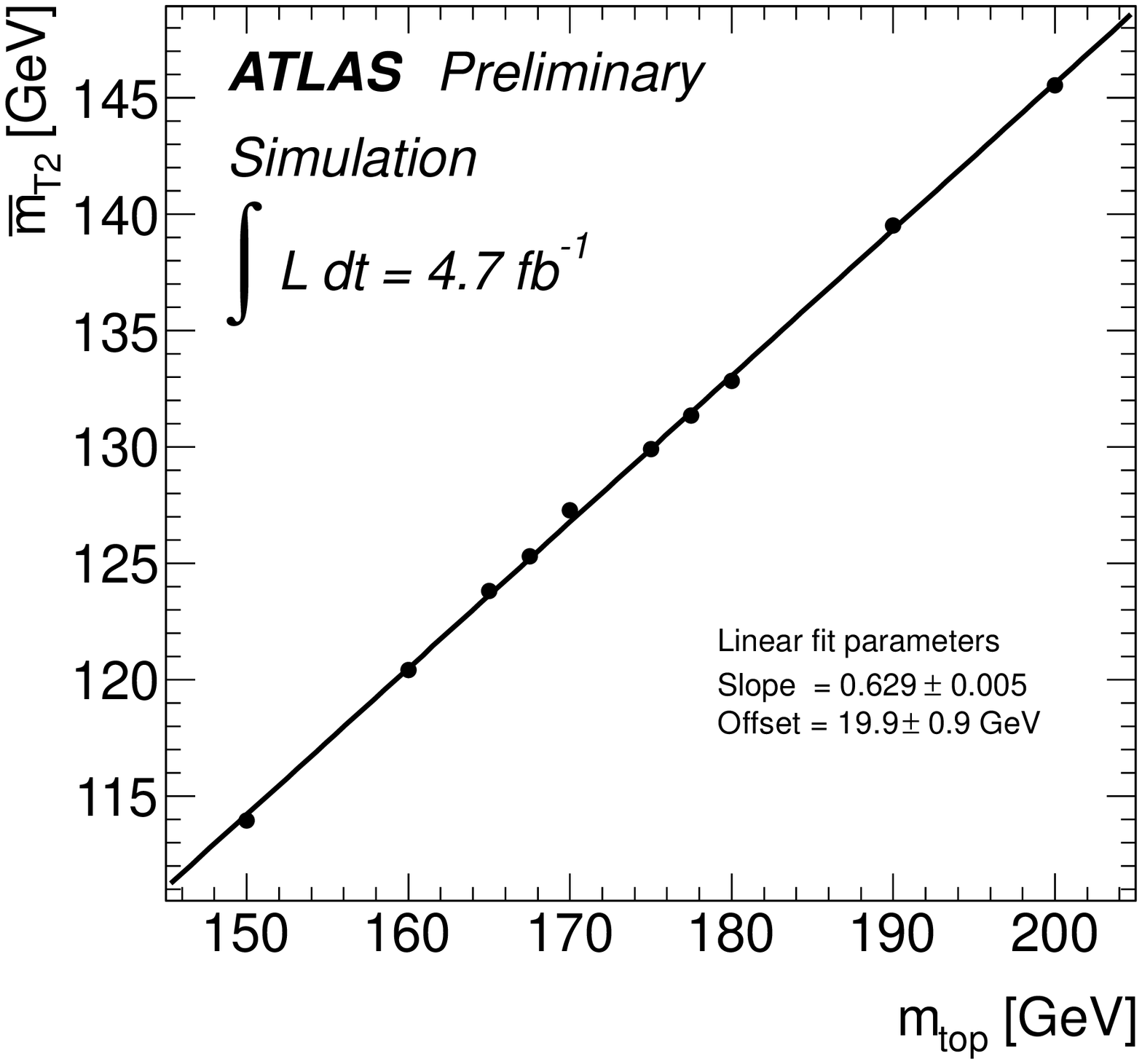}
  \end{subfigure}
  \caption{Left: distribution of $m_{T2}$ on the selected events, for collision data and simulation. Right: Calibration curve based on Monte Carlo simulation of $t\bar{t}$ events at different input top quark masses including all expected backgrounds. The uncertainties are statistical only \cite{bib:calibc}. \label{fig:calibc}}
\end{figure}

The value of $m_{T2}$ is directly related to $m_{top}$: by means of a calibration curve built on MC, the latter can be inferred from the former on collision data. While a distribution of values for $m_{T2}$ is populated using all the selected events, the mean of such distribution is used to extract the reference value used in the calibration curve. The distributions of $m_{T2}$ in data and the calibration curve are shown in Figure \ref{fig:calibc}.  

The results obtained with this technique yield $m_{top}$ = 175.2 $\pm$ 1.6 (stat) $^{+3.1}_{-2.8}$ (syst) GeV. The dominant systematic uncertainties are from the $t\bar{t}$ generator (1.3 GeV), parton showering (0.9 GeV) and colour reconnection (1.2 GeV) models, which all affect the final state kinematics; and a residual dependence on JES ($^{+1.6}_{-1.4}$ GeV) and b-JES ($^{+1.5}_{-1.2}$ GeV) from the event selection. 

\section{Summary}
In summary, a set of direct top quark mass measurements is available from the ATLAS experiment at the LHC obtained with two main techniques (template and calibration curve) from all the main channels. At the moment, the most precise measurement (from a two-dimensional template technique on the lepton+jets signature) attains an uncertainty of 2.4 GeV in total, about 1.4$\%$ relative to the measured mass value of 174.5 GeV. On-going work is especially aimed at constraining the MC model varations with data so to obtain a more realistic estimate of the systematic uncertainties. An indirect measurement from the $t\bar{t}$ cross-section was performed in 2010 with 35 pb$^{-1}$ of integrated luminosity: $m_{top}$ = 166.4 $^{+7.8}_{-7.3}$ (syst) GeV \cite{bib:fromxsec}.

\end{document}